\PassOptionsToPackage{pdfpagelabels=false}{hyperref}
\documentclass[letters,fleqn,usenatbib]{mnras}

\usepackage{txfonts}

\usepackage[T1]{fontenc}
\usepackage{ae,aecompl}

\usepackage{graphicx}	% Including figure files
\usepackage{tablefootnote}
\title[]{The proper motion of HV2112: A T\.{Z}O candidate in the SMC}

\author[C.~C. Worley et al.]{C.~Clare Worley,\thanks{E-mail: ccworley@ast.cam.ac.uk}
Mike.~J. Irwin,
Christopher~A. Tout,
Anna~N. \.{Z}ytkow,
\newauthor
 Morgan Fraser
and Robert.~G. Izzard
\\
Institute of Astronomy, University of Cambridge, Madingley Rise, Cambridge CB3 0HA, U.K.
}

\date{Accepted 2016 February 26. Received 2016 February 25; in original form 2016 February 15}

\pubyear{2016}

\begin{document}
\label{firstpage}
\pagerange{\pageref{firstpage}--\pageref{lastpage}}
\maketitle

\begin{abstract}
The candidate Thorne--\.{Z}ytkow object (T\.{Z}O),  HV2112, is becoming a well-studied if enigmatic object. A key point of its candidacy as a T\.{Z}O is whether or not it resides in the Small Magellanic Cloud (SMC). HV2112 has detections in a series of photometric catalogues which have resulted in contradictory estimates of its proper motion and, therefore, its membership within the SMC. This letter seeks to resolve the issue of the SMC membership of HV2112 through a reanalysis of extant photometric data. We also demonstrate the difficulties and downfalls inherent in considering a range of catalogue proper motions. We conclude that the proper motion, and associated ancillary radial velocity, positional and photometric properties, are fully consistent with HV2112 being within the SMC and thus it remains a candidate T\.{Z}O. 
\end{abstract}


%%%%%%%%%%%%%%%%%%%%%%%%%%%%%%%%%%%%%%%%%%%%%%%%%%

%%%%%%%%%%%%%%%%% BODY OF PAPER %%%%%%%%%%%%%%%%%%

\section{Introduction}
HV2112 has recently been proposed \citep{Levesque2014} as a likely candidate for a Thorne--\.{Z}ytkow object (T\.{Z}O), a red super giant with a neutron star core \citep{Thorne1975,Thorne1977}. This candidacy depends on HV2112 being a member of the Small Magellanic Cloud (SMC). \citet{Maccarone2016} propose an estimate of a proper motion (PM) for HV2112 which, if in the SMC, corresponds to a space motion of $3000\, \rm km \,s^{-1}$, exceeding the escape velocity of SMC. The PM of \citet{Maccarone2016} implies a reasonable assumption of HV2112 being a Milky Way halo star at a distance of 3\,kpc. Residence in the halo, at a closer distance by a factor of ten or so, would mean that HV2112 is not sufficiently luminous to be a red super giant, let alone a T\.{Z}O.

HV2112 has also been found to have a strong calcium line in its spectrum \citep{Levesque2014}. Calcium is potentially a key discriminator between the proposed sites of origin for this star. However, the detected calcium is more in line with levels expected for halo stars, rather than the SMC. If, as we support here, HV2112 is indeed a luminous SMC giant, the strong calcium line may well be key to understanding its evolution \citep{Tout2014,Sabach2015}.

\section{The Proper Motion of HV2112}
A range of photometric catalogues contain images of HV2112. \citet{Maccarone2016} investigated the PM of HV2112 from the Southern Proper Motion (SPM) survey \citep{Girard2011}. Observations were made in two different epochs, the first in the B band in 1972 and the second in the V band in 2007, providing a 35\,yr baseline between epochs. A PM of $2.8\pm2.3 \, \rm mas\,yr^{-1}$ in right ascension and $-9.8\pm2.3 \, \rm mas\,yr^{-1}$ in declination was obtained from the SPM catalogue indicating a space motion of $3000\, \rm km \,s^{-1}$ if HV2112 is an SMC member. As noted by \citet{Maccarone2016} there is a significant discrepancy in the direction of declination between the SPM PM and that provided in the UCAC4 catalogue \citep{Zacharias2013}. The UCAC4 PM estimate is $1.8\pm2.9 \, \rm mas\,yr^{-1}$ in right ascension and $-3.3\pm2.7 \, \rm mas\,yr^{-1}$ in declination. The available literature PMs are explored more in Section~\ref{sec:litrev}.

To further investigate the PM of HV2112 we made two additional independent studies. The first compared images in the R-band from a UK Schmidt sky survey plate \citep{Cannon1975}, taken in 1989, to images in the near-infra red (NIR) Y-band from VISTA \citep{Emerson2004} taken in 2012.  Secondly NIR J-band images from VISTA, also from 2012, were directly compared with the 2MASS Point Source Catalog \citep[PSC,][]{Skrutskie2006} of the same region taken in 1998.

Both sets of data were used to generate PM estimates for HV2112 and the surrounding field stars. This was limited to a 5\,arcmin $\times$ 5\,arcmin region centred on HV2112, for the photographic plate -- VISTA comparison, to minimise the effects of differential refraction given the different passbands used. For the VISTA -- 2MASS comparison, a larger region approximately 1\,degree $\times$ 1\,degree in size could be used given the similar NIR passbands.

The photographic plate catalogue was directly matched to the VISTA Y-band catalogue with a six-constant linear mapping in standard coordinates ($\xi, \eta$) with tangent point at the nominal position of HV2112. The direct 
match between the photographic plate and the VISTA NIR data benefits from the vastly increased number of objects detected (599 and 1550 to R and Y with limiting magnitudes of approximately 19 and 20 respectively), compared to the 72 suitable 2MASS PSC sources in the 5 arcmin $\times$ 5 arcmin region. 

\begin{figure}
	\centering
\includegraphics[width=70mm,angle=-90]{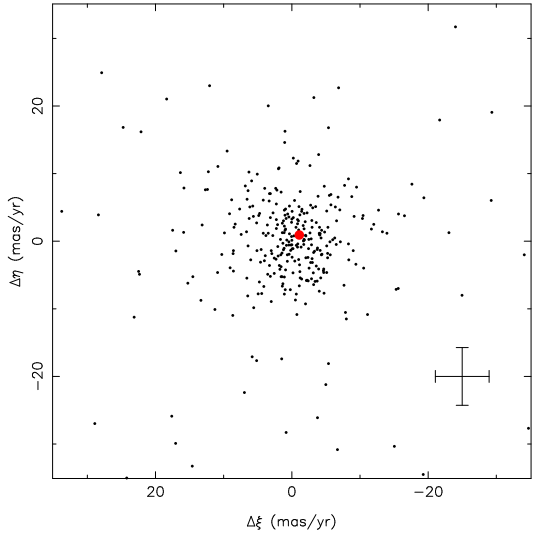}
\caption{Derived PM from UK Schmidt R Band (1989) and VISTA Y (2012) for SMC 
field stars in black and HV2112 in red. Standard coordinate notation is used where $\xi$ and $\eta$ are in the RA and Dec orientations respectively. Cross hairs are derived from the measurement errors.}
\label{fig:tzo_pm_uk}
\end{figure}

In this relatively deep data SMC stars dominate the field population. Therefore the effective PM reference frame is defined by the mean heliocentric PM of the SMC. The measured PM for HV2112 based on the resulting 23\,yr baseline is $-1.09\pm4.27\, \rm mas\,yr^{-1}$ in right ascension and $0.92\pm4.40\, \rm mas\,yr^{-1}$ in declination as shown in Fig.~\ref{fig:tzo_pm_uk}. The errors here are dominated by the photographic plate $rms$ error. For well-measured stars like HV2112, the $rms$ is typically 100\,mas, corresponding to $4.3\, \rm mas\,yr^{-1}$ over the 23\,yr baseline. This is on the order of the scatter in the SMC field in Fig.~\ref{fig:tzo_pm_uk}.

\begin{figure}
	\centering
\includegraphics[width=70mm]{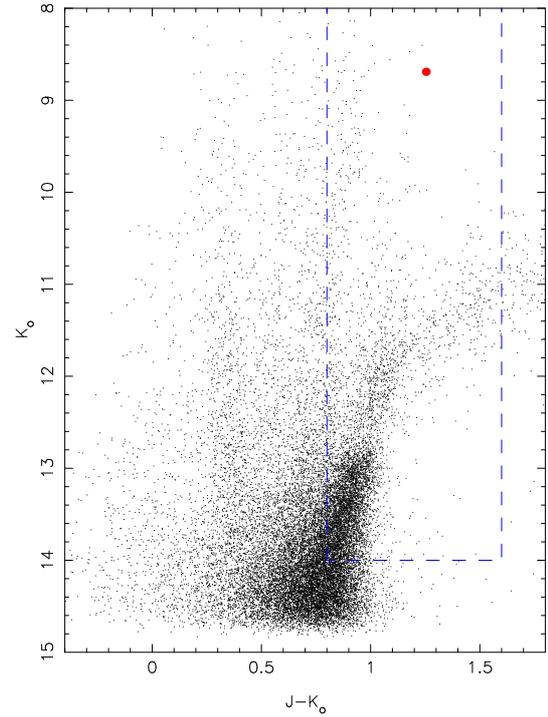}
\caption{Extinction-corrected colour--magnitude diagram from 2MASS for all point sources (black) lying within 1 degree of HV2112 (red). The blue box selects a field dominated by SMC member stars, rejecting blueward foreground dwarfs.}
\label{fig:tzo_cmd}
\end{figure}

\begin{figure}
	\centering
\includegraphics[width=70mm,angle=-90]{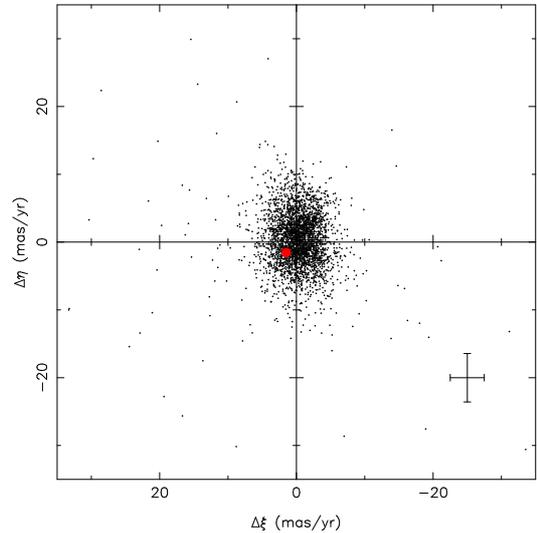}
\caption{Derived PM from 2MASS (1989) and VISTA J (2012) for SMC field stars in black, selected from within the blue box in Fig~\ref{fig:tzo_cmd}, and HV2112 in red. Cross hairs indicate the measurement errors which are dominated by the 2MASS position.}
\label{fig:tzo_prop}
\end{figure}

The VISTA catalogues are astrometrically calibrated with 2MASS stars for each pointing. This enables direct PM measurements.  The procedure is illustrated in Figs~\ref{fig:tzo_cmd} and ~\ref{fig:tzo_prop}.
Fig.~\ref{fig:tzo_cmd} shows an extinction-corrected 2MASS colour--magnitude diagram for the region.  The selection box for the PM estimates shown in Fig.~\ref{fig:tzo_prop} is highlighted by the blue dashed lines and the location of HV2112 shown in red. The giant and supergiant populations of the SMC are prominent and HV2112 sits (notably) at the top of the M-supergiant locus. The selection box has a two-fold purpose. First it ensures that SMC stars dominate the PM collection (rejecting the blueward foreground dwarfs) and secondly limits the 2MASS stars used to those with the lowest $rms$ positional errors \citep[see for example fig.~20 of][]{Skrutskie2006}. The SMC field stars cluster tightly near the origin in Fig.~\ref{fig:tzo_prop} and the PM of HV2112 is highlighted in red.

The reference frame is again defined by the mean heliocentric PM of the SMC, in this case over a 14\,yr baseline and yields a PM for HV2112 of $1.48\pm2.49\, \rm mas\,yr^{-1}$ in right ascension and $-1.55\pm3.57\, \rm mas\,yr^{-1}$ in declination.  The errors here are dominated by the 2MASS positional uncertainties which are consistent with the $rms$ errors derived from the locus of SMC points in the figure. Proper motion estimates for HV2112 from several independent 2012 VISTA measurements show a negligible scatter of $\pm0.1 \rm mas\,yr^{-1}$.

\section{Literature Proper Motions}\label{sec:litrev}
As discussed in this Letter, the association of HV2112 with the SMC depends largely on the measurement and interpretation of its PM. We have derived an accurate PM for HV2112 based on a re-analysis of the best available imaging data. However, were our analysis not available, it would be necessary to resort to published catalogue proper motions. In the following section, we consider such an approach for HV2112.

We have reviewed all HV2112 PMs available through VizieR\footnote{http://vizier.u-strasbg.fr/viz-bin/VizieR} \citep{Ochsenbein2000}. The catalogues and associated PMs are listed in Table~\ref{tab:literature}. Column $r$ is the coordinate distance between the catalogue and SIMBAD for HV2112. The PMs accepted for the SMC are $0.772\pm0.063\, \rm mas\,yr^{-1}$ in RA and $-1.117\pm0.061\, \rm mas\,yr^{-1}$ in Dec \citep{Kallivayalil2013}. 

No error columns are provided for the XPM catalogue but the catalogue description provides an estimate of the random errors of the absolute PMs for southern hemisphere objects of between $5$ and $10\, \rm mas\,yr^{-1}$. Thus the errors were assumed to be $5\, \rm mas\,yr^{-1}$ for both measurements.

The proper motion calculated by \citet{Maccarone2016} as a weighted mean of the SPM and UCAC4 PMs is also included in Table~\ref{tab:literature}. However, as it is a combination of two of the catalogue PMs, we do not include it in the comparison of the literature PMs.

Five entries (indicated by {\color{red}*} and shown in red in Table~\ref{tab:literature}) are discarded from the discussion for the following reasons:
\begin{itemize}
 \item NOMAD is a duplicate of UCAC2;
 \item UCAC2 has been superceded by UCAC4;
 \item PPMX PM has a large offset in coordinate distance ($r=1.4112'$);
 \item IGSL is a duplicate of UCAC4;
 \item AllWISE PM has excessive errors.
\end{itemize}

\begin{table}
\caption{Literature PMs for HV2112 taken from catalogues listed in VizieR plus the \citet{Maccarone2016} (M\&dM2016) weighted mean PM and the two measurements presented here, IoA:UKV (UKSchmidt$+$VISTA) and IoA:2MV (2MASS$+$VISTA).}
\tabcolsep=0.09cm
\begin{tabular}{lccccc}
\hline\hline 
Catalogue & $r$ & PM RA & $\sigma_{\rm PMRA}$ & PM Dec & $\sigma_{\rm PMDec}$ \\
          & (') & $(\rm mas\,yr^{-1})$ & $(\rm mas\,yr^{-1})$ & $(\rm mas\,yr^{-1})$ & $(\rm mas\,yr^{-1})$ \\
\hline
{\color{red} *NOMAD$^a$}   & 0.0004 & 8.00 & 6.30 & -8.10 & 5.90 \\ 
{\color{red} *UCAC2$^b$} & 0.0004 & 8.00 & 6.30 & -8.10 & 5.90 \\ 
{\color{red} *PPMX$^c$}   & {\color{red} 1.4112} & 8.78 & 20.10 & 4.34 & 20.10\\ 
PPMXL$^d$ & 0.0013 & 14.40 & 14.20 & -1.30 & 14.20 \\ 
XPM(PSC)$^e$ & 0.0003 & 10.94 & 5.00 & 5.36 & 5.00\\ 
XPM(XSC)$^e$ & 0.0003 & 9.14 & 5.00 & 4.42 & 5.00 \\ 
SPM$^f$   & 0.0027 & 2.80 & 2.27 & -9.78 & 2.30\\ 
UCAC4$^g$   & 0.0003 & 1.80 & 2.90 & -3.30 & 2.70 \\ 
{\color{red} *IGSL$^h$}   & 0.0003 & 1.80 & 2.90 & -3.30 & 2.70\\ 
APOP$^i$ & 0.0018 & 0.00 & 10.10 & 2.60 & 0.40\\ 
{\color{red} *AllWISE$^j$}   & 0.0005 & -2.00 & {\color{red} 30.00} & 28.00 & {\color{red} 33.00}\\ 
IoA:UKV & 0.0040 & -1.09 & 4.27 & 0.92 & 4.40\\ 
IoA:2MV & 0.0010 & 1.48 & 2.49 & -1.55 & 3.57\\ 
M\&dM2016$^\#$ & & 2.40 & 1.80 & -6.80 & 1.80 \\ 
\hline
\end{tabular}
\label{tab:literature}
a.~\citet{Zacharias2004a}; b.~\citet{Zacharias2004b}; c.~\citet{Roser2008}; d.~\citet{Roeser2010}; e.~\citet{Fedorov2011}; f.~\citet{Girard2011}; g.~\citet{Zacharias2013}; i.~\citet{Smart2014}; i.~\citet{Qi2015}; j.~\citet{Wright2010}.\\
$^\#$ Weighted mean of the SPM and UCAC4 PMs. \\
{\color{red}* Discarded from the literature comparison.}
\end{table}

\begin{figure*}
\centering
\begin{minipage}{175mm}
%\hspace{-2.0cm}
\includegraphics[width=180mm]{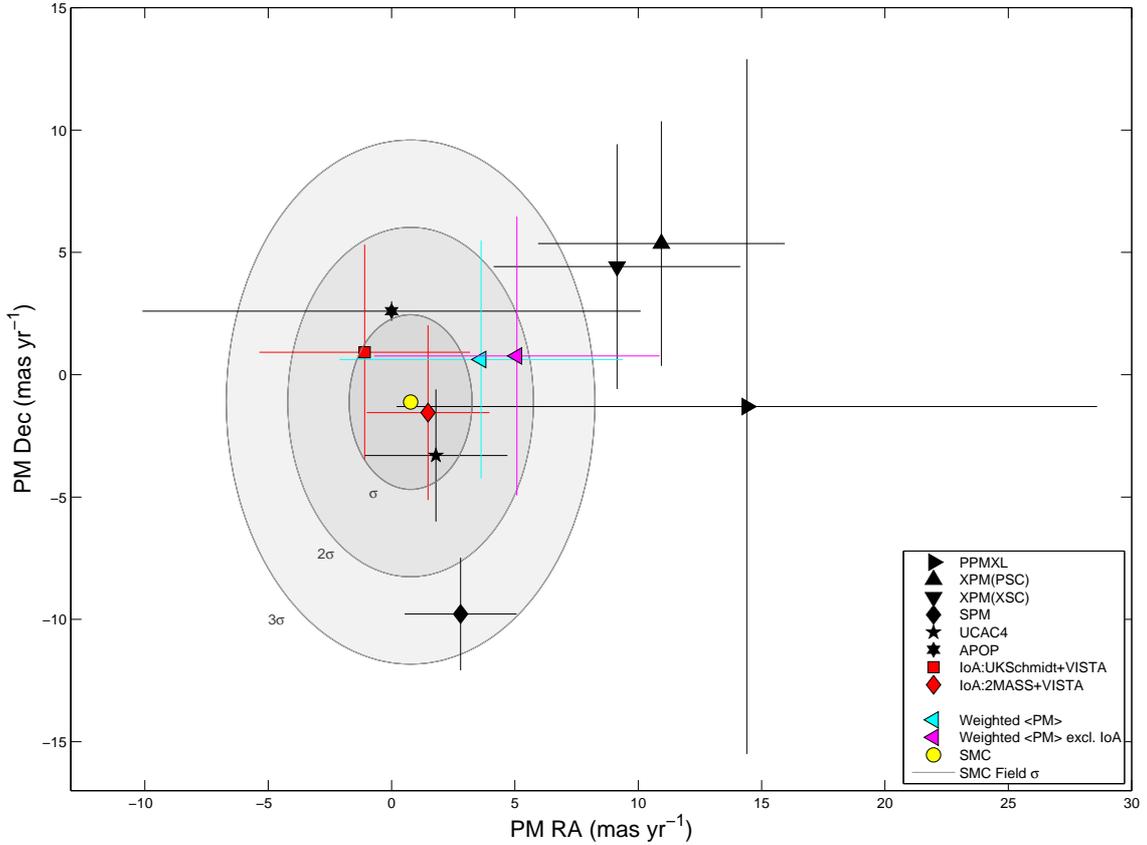}
\vspace{-1.0cm}
%\caption{a) All literature PMs listed in VizieR, plus the IoA measurements presented here, plus the Mean and Weighted Mean of the HV2112 measurements with and without the IoA PMs, plus the SMC PM. b) As for a) but with 3 catalogues results removed as outlined in the text.}\label{fig:literature}
\caption{HV2112 literature PMs selected from VizieR, plus the IoA measurements presented here, plus the weighted mean of the literature measurements (with and without the IoA PMs), and the SMC PM. The SMC field PM distribution is also shown as a series of ellipses. The semi-minor and semi-major axes are integer multiples of the IoA:2MASS$+$VISTA RA and Dec PM uncertainties respectively. IoA:2MASS$+$VISTA RA and Dec PM uncertainties represent the uncertainty ($\sigma$) in the SMC field PM distribution as shown in Fig.~\ref{fig:tzo_prop}}\label{fig:literature}
\end{minipage}
\end{figure*}

Fig.~\ref{fig:literature} displays the remaining six HV2112 literature PMs with those presented here (IoA). The weighted mean of the literature PMs with and without the IoA PMs are also displayed, as well as the accepted SMC PM. Also shown are a series of three ellipses, where the semi-minor and semi-major axes are integer multiples of the IoA:2MASS$+$VISTA RA and DEC PM uncertainties respectively. These uncertainties are derived from the SMC field as described above and illustrated in Fig.~\ref{fig:tzo_prop} and therefore can be considered as the uncertainty ($\sigma$) in the SMC field PM distribution.

%\begin{figure}
%\centering
%\begin{minipage}{85mm}
%\hspace{-1.0cm}
%\includegraphics[width=105mm]{/home/cclare/Stars/TZO/ProperMotion/PropMotPap/LIteraturePM/litpm_contours.eps}
%\vspace{-1.0cm}
%\caption{a) All literature PMs listed in VizieR, plus the IoA measurements presented here, plus the Mean and Weighted Mean of the HV2112 measurements with and without the IoA PMs, plus the SMC PM. b) As for a) but with 3 %catalogues results removed as outlined in the text.}\label{fig:literature}
%\end{minipage}
%\end{figure}

One option for us is to define a PM for HV2112 by taking a weighted mean of the literature PMs, accounting for the range in the magnitudes of the associated PM errors. Excluding the IoA results from the weighted mean causes us a small shift in the RA direction as shown in Fig.~\ref{fig:literature}. But both weighted means are within 1$\,\sigma$ of the SMC PM within their errors. When we consider the literature PMs and IoA PMs together, five agree with the SMC PM to 1$\,\sigma$ within their errors. Two agree to 2$\,\sigma$ and one agrees to 3$\,\sigma$. 

However, making a simple comparison between literature PMs, as above, ignores whether or not the PM reference frames are consistent and therefore comparable. Certainly calculating a mean PM is not valid if the reference frames are not consistent. 

As noted above, the IoA measurements are basically heliocentric with respect to the SMC PM. For the two measurements used by \citet{Maccarone2016}, the PMs determined for the SPM catalogue use galaxies to establish a PM zeropoint. Likewise UCAC4 is based on the Tycho2 ICRS \citep{Hog2000} linkage and so is also zeropointed with an extragalactic reference frame.  Thus these two PMs are also heliocentric and are based on an extragalactic reference frame.

Comparing these four PMs, the two results that agree most are UCAC4 and IoA:2MASS$+$VISTA to 0.1$\,\sigma_{\rm PMRA}$ and 0.4$\,\sigma_{\rm PMDec}$. Here $\sigma_{\rm PMRA}$ and $\sigma_{\rm PMDec}$ are the respective PM errors summed in quadrature. SPM and UCAC4 agree to 0.3$\,\sigma_{\rm PMRA}$ and 1.8$\,\sigma_{\rm PMDec}$. The greater disagreement in the PM in declination is evident in Fig.~\ref{fig:literature} where the SPM PM in declination is a clear outlier. 

When comparing to the SMC directly, as shown in Fig.~\ref{fig:literature}, both UCAC4 and the IoA PMs are in good agreement with the SMC PM within 1$\,\sigma$. Also the SPM PM agrees with the SMC PM to 2$\,\sigma$. Thus Fig.~\ref{fig:literature} shows that the literature PMs are generally consistent with the SMC PM although their distribution is quite scattered.

It is clear that several of the catalogues have PMs for HV2112 which are in disagreement by more than their quoted uncertainties. Furthermore, it can be unclear which catalogues to include in a comparison and one must be wary of cherry-picking the data by rejecting unfavourable measurements. We also note that many catalogues rely on overlapping data sets (e.g. SPM and UCAC4 share a common first epoch from SPM), and so these should not be considered as independent measurements of the true PM. 

In light of such issues, HV2112 provides an excellent example of the potential pitfalls associated with extracting PMs from the literature. We argue that the new measurements we present here are the best PM measurements to-date for HV2112. In the very near future positions, and later PMs and parallaxes, will be available from the Gaia Mission \citep{Perryman2001}. These will provide the definitive answer on the true location of HV2112.

\section{Discussion}
The two PM analyses carried out here strongly suggest that HV2112 is a member of the SMC. In this study, as shown in Fig.~\ref{fig:tzo_prop}, HV2112 is located well within the cluster of SMC points whereas the PM proposed by \citet{Maccarone2016} would put HV2112 outside of the SMC field population.

The reflex solar PM for a stationary halo star at 3\,kpc would be $-8.94\, \rm mas\,yr^{-1}$ in RA and $9.21\, \rm mas\,yr^{-1}$ in Dec. However for a halo star at 3\,kpc a high transverse motion is expected and so the \citet{Maccarone2016} PM would not be unreasonable.

\begin{figure}
\centering
\includegraphics[width=82mm]{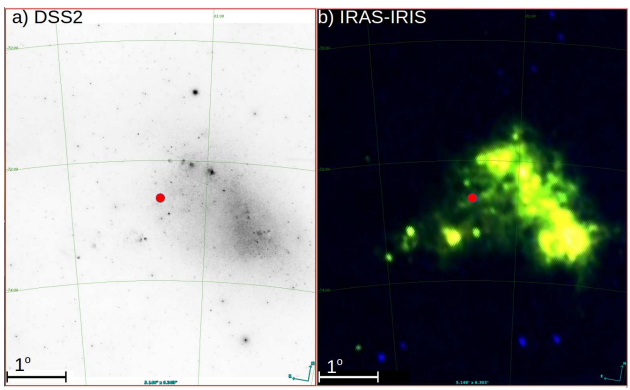}
\caption{Location of HV2112 (red circle) in the SMC generated using Aladin \citep{Bonnarel2000} with a) DSS2; b) IRAS-IRIS \citep{Miville2005} images. HV2112 not only lies in the direction of the SMC but appears close to a region of relatively recent star formation.}
\label{fig:hv2112inSMCir}
\end{figure}

Other types of measurements should be considered alongside the PM determination to provide a broader picture. For example, the difference in RA and Dec of HV2112 from the SMC positional centroid is $\Delta \rm RA = 261.5\,'$ and $\Delta \rm Dec = 11.1\,'$. While located in the outer edges of the angular extent of the SMC ($\rm major\,\,axis = 309.0\,'$ and $\rm minor\,\,axis = 204.1\,'$) as shown Fig.~\ref{fig:hv2112inSMCir}, HV2112 lies coincident with the substructure of the east wing of the SMC. The east wing is evidence of star forming events that occurred between $50 - 200\,$Myr ago \citep{Irwin1990} and is populated by young massive stars akin to HV2112.

The measured radial velocity of HV2112 \citep[approximately $157\, \rm km \,s^{-1}$,][]{Levesque2014} is in good agreement with the accepted radial velocity of the SMC \citep[$145.6\, \rm km \,s^{-1}$,][]{McConnachie2012}. The Galactocentric line-of-sight radial velocity for HV2112 of approximately $13\, \rm km \,s^{-1}$ is consistent with both a halo star and membership of the SMC. However the velocity dispersion of the SMC is narrower \citep[$\sigma_\mathrm{SMC} = 27\, \rm km \,s^{-1}$,][]{Harris2006} than that of the halo \citep[$\sigma_\mathrm{halo} > 85\, \rm km \,s^{-1}$,][]{Brown2010}. Both encompass HV2112 thus favouring membership of neither population in particular.

Finally, in the 2MASS colour--magnitude diagram of point sources lying within 1\,degree of HV2112 shown in Fig.~\ref{fig:tzo_cmd}, HV2112 lies clearly on the SMC M supergiant locus. If HV2112 is a halo star at 3\,kpc, that it has an absolute magnitude at the 2MASS epoch which places it exactly on the SMC supergiant locus would be intriguingly coincidental.

When considered in combination, that an object has coordinate position, PM and radial velocity in good agreement with the SMC and has photometry placing it clearly on the SMC supergiant locus, is strong evidence for HV2112 being a member of the SMC. To have all these in agreement but to not be an SMC member seems unlikely.

\section{Conclusion}
This Letter summarises independent analyses of the PM of HV2112. These PM analyses as well as the coordinate position, radial velocity and photometric measurements of HV2112 are all consistent with and strongly support the assumption that HV2112 is a member of the SMC. Therefore HV2112 is not excluded as a candidate T\.{Z}O or a luminous, super-AGB star \citep{Tout2014}.

The study of HV2112 is ongoing with high and medium resolution spectroscopic observations. Spectral energy distribution and chemical abundance analyses may reveal the crucial characteristics that can discriminate between the various proposed origins of this enigmatic star.

\section*{Acknowledgements}
This work is based on observations collected at the European Organisation for Astronomical Research in the Southern Hemisphere under ESO programme(s) 179.B-2003 and was partly supported by the European Union FP7 programme through ERC grant number 320360. RGI thanks the STFC for funding for his Rutherford fellowship. CAT thanks Churchill College for his fellowship.

This research has made use of the the VizieR catalogue access tool \citep{Ochsenbein2000}, the Aladin sky atlas and the SIMBAD database developed and operated at CDS, Strasbourg, France. The Digitized Sky Surveys (DSS) were produced at the Space Telescope Science Institute under U.S. Government grant NAG W-2166. The images of these surveys are based on photographic data obtained using the Oschin Schmidt Telescope on Palomar Mountain and the UK Schmidt Telescope. The plates were processed into the present compressed digital form with the permission of these institutions.

\vspace{-0.5cm}

%%%%%%%%%%%%%%%%%%%%%%%%%%%%%%%%%%%%%%%%%%%%%%%%%%


%\bibliographystyle{mnras}
%Included for Gather Purpose only:
%input "\home\cclare\Stars\TZO\ProperMotion\PropMotPap\astroph\propmot_tzo.bib"
%\setlinespacing{1.44}
%\bibliography{propmot_tzo}

%\bibliographystyle{mnras}

% Don't change these lines
\bsp	% typesetting comment
\label{lastpage}
\end{document}